\documentclass[twocolumn,showpacs,preprintnumbers,amsmath,amssymb]{revtex4}

\usepackage{graphicx}

\usepackage{dcolumn}

\usepackage{bm}

\begin{document}

\preprint{APS/123-QED}

\title{Role of Single Defects in Electronic Transport \\ through Carbon
Nanotube Field-Effect Transistors}

\author{Marcus Freitag}

\affiliation{Department of Physics and Astronomy and Laboratory for
Research on the Structure of Matter, University of Pennsylvania, 209
South $33^{rd}$ St., Philadelphia, PA 19104.}

\author{Sergei V. Kalinin}

\affiliation{Department of Materials Science and Engineering,
University of Pennsylvania, 3231 Walnut St., Philadelphia, PA  19104.}

\author{Dawn A. Bonnell}

\affiliation{Department of Materials Science and Engineering,
University of Pennsylvania, 3231 Walnut St., Philadelphia, PA  19104.}

\author{A.T. Johnson$^*$}

\affiliation{Department of Physics and Astronomy and Laboratory for
Research on the Structure of Matter, University of Pennsylvania, 209
South $33^{rd}$ St., Philadelphia, PA 19104.}

\date{\today}

\begin{abstract}

The influence of defects on electron transport in single-wall carbon
nanotube field effect transistors (CNFETs) is probed by combined
scanning gate microscopy (SGM) and scanning impedance microscopy
(SIM). SGM reveals a localized field effect at discrete defects along
the CNFET length. The depletion surface potential of individual
defects is quantified from the SGM-imaged radius of the defect as a
function of tip bias voltage. This provides a measure of the Fermi
level at the defect with zero tip voltage, which is as small as 20
meV for the strongest defects. The effect of defects on transport is
probed by SIM as a function of backgate and tip-gate voltage. When
the backgate voltage is set so the CNFET is ``on" (conducting), SIM
reveals a uniform potential drop along its length, consistent with
diffusive transport. In contrast, when the CNFET is ``off", potential
steps develop at the position of depleted defects. Finally,
high-resolution imaging of a second set of weak defects is achieved
in a new ``tip-gated" SIM mode.
\end{abstract}





\maketitle

Unique structural and electronic properties of single-wall carbon
nanotubes (SWNTs) allow them to act as molecular wires \cite{Tans1}
and switching elements in nanoscale devices and logic
gates.\cite{Tans2,Martel1,Derycke,Bachtold1,Radosavljevic,Fuhrer1}
Successful implementation of SWNT electronic devices necessitates
quantitative characterization of local structure and properties.
Scanning tunneling microscopy resolves local atomic and electronic
structure,\cite{Wildoer} but application to nanotube circuits is
problematic because tip current feedback cannot be used when the tip
is over an insulating substrate. Local characterization of {\it
active devices} can be done with recently developed techniques such
as scanning gate microscopy (SGM),\cite{Bachtold2,Tans3}
electrostatic force microscopy,\cite{Bachtold2} and scanning
impedance microscopy (SIM).\cite{Kalinin}

Schottky barriers form where semiconducting tubes contact metal
electrodes\cite{Freitag1,Martel2} or metallic tubes.\cite{Fuhrer2}
These barriers can be imaged by SGM and controlled electrostatically
by a voltage-biased SGM tip.\cite{Freitag1,Freitag2} SGM can also
resolve valance\cite{Bachtold2,Tans3} and
conduction\cite{Radosavljevic} band potential modulations in carbon
nanotube field-effect transistors (CNFETs). These might be associated
with atomic defects on the SWNT, e.g., bond-rotations, add-dimers, or
vacancies. Alternatively, they might be ascribed to dopants such as
adsorbed or encapsulated impurities,\cite{Hornbaker} or trapped
charges in the substrate.

Here we present combined SGM and SIM measurements that illuminate the
role of defects on electron transport in CNFETs. Through analysis of
electrostatic interactions in SGM, we quantify the depletion surface
potential for single defects and the size of their associated valence
band modulations, key inputs to the design of CNFET devices. SIM is
used to measure the potential distribution in the CNFET at different
back gate voltages. We observe a crossover from diffusive conduction
along the full length of the nanotube to conduction inhibited at
barriers localized at depleted defects. Finally, we present a SIM
mode where the tip simultaneously perturbs the local density and
probes the electrostatic potential. This ``tip-gating" mode of SIM
reveals a set of weak defects that are not resolved in conventional
SGM.

Carbon nanotubes are grown by catalytic chemical vapor
deposition\cite{Kong} on a SiO2/Si substrate. SWNTs are identified by
an apparent height less than 3 nm in AFM and then contacted by Cr/Au
electrodes defined by e-beam lithography. Semiconducting nanotube
samples display a strong field effect in response to a back-gate
voltage applied to the degenerately-doped silicon substrate. Figure
1(a) shows the SGM/SIM measurement set-up, based on a Digital
Instruments Dimension 3000 NS-IIIA AFM using gold-coated tips
(CSC12A, Micromasch). Interleave scans are used for SGM and SIM with
a lift height of about 10 nm. CNFET bias voltage $V_{ac}$ is $0.1-1$
$V_{pp}$ at a drive frequency a few kilohertz away from the
cantilever resonance. To reduce screening due to charge injection
from the tube into oxide traps,\cite{Radosavljevic,Fuhrer1} the back
gate is triggered by the microscope line-synchronization signal.

In SGM, a conductive tip with an applied potential acts as a
spatially localized gate scanned under AFM-feedback near the
voltage-biased CNFET. The image formed from the transport current as
a function of tip position reveals precise locations where the CNFET
has a strong response to the tip gate. We use AC-SGM, with an AC
modulation applied to the sample and current measured with a lock-in
amplifier, significantly reducing measurement noise compared to the
DC mode used in earlier work.

Our samples are p-type CNFETs, so the Schottky barrier at the
positively biased electrode is under reverse bias and limits the
current. A negatively-biased tip positioned near the reverse biased
Schottky barrier gives marked current enhancement,\cite{Freitag1}
showing that the barrier can be lowered
electrostatically.\cite{Freitag1,Martel2} In DC-SGM, only the barrier
at the positive electrode is imaged. In AC-SGM, each electrode is at
positive bias during half of the AC cycle, and {\it both} Schottky
barriers are imaged (data not shown). At positive bias, the tip
probes valence band modulations by depleting carriers (holes)
locally. As described in the following paragraphs, images in
``depletion mode SGM" are used to quantify the depletion surface
potential and Fermi level associated with individual defects.

SGM images at different tip voltages are shown in Figs. 2(a)-(e).
Four strong defects are visible, labeled 1 - 4. The imaged diameter
of each defect increases linearly with tip voltage (Fig. 3). We explain
this observation by considering the surface potential distribution
below the tip using image charges appropriate for a spherical tip
above a dielectric.\cite{Jackson} This approach is applicable for
small tip-surface separation, $d \ll R$, where $R$ is the tip radius.
We assume that the surface potential is not screened by the nanotube;
this is expected if the Debye length in the CNFET is significantly
larger than the tube diameter, e.g., when the neighborhood of a
defect is nearly depleted. The surface potential below the tip then
becomes:

\begin{equation} V(\rho )=\frac{1}{4\pi \epsilon_{0}}
\frac{2}{\kappa+1}\sum_{i=0}^{\infty}
\frac{Q_{i}}{\sqrt{(R+d-r_{i})^{2}+\rho^{2}}} \end{equation}

where $\rho$ is the distance from tip projection on the surface and
$Q_{i}$ a set of image charges located distances $r_{i}$ from the
center of the sphere:

\begin{equation}
Q_{i+1}=\frac{\kappa-1}{\kappa+1}\frac{R}{2(R+d)-r_{i}}Q_{i},
\end{equation}

\begin{equation} r_{i+1}=\frac{R^{2}}{2(R+d)-r_{i}} \end{equation}

$R$ is the tip radius, $d$ the tip-surface separation, $Q_{0}=4\pi
\epsilon_{0}RV_{tip}$, $r_{0}=0$, $\kappa=3.9$ the dielectric
constant of $SiO_{2}$, and $V_{tip}$ the tip bias. For $\rho \gg d,R$
the surface potential becomes
$V(\rho)=2 \alpha R V_{tip}/(\kappa +1)\rho$, where $\alpha$ is the
ratio of the tip-surface capacitance to that of a sphere of radius
$R$.\cite{Kalinin3}  Current through the CNFET is suppressed at a
defect if the local surface potential is $V \ge V^{*}$, where $V^{*}$
is the defect depletion surface potential.

The SGM-imaged diameter of the $i$-th defect therefore increases
linearly with $V_{tip}$ (Fig. 3): $D_{i}=(V_{tip}/V^{*}_{i})4\alpha
R/(\kappa +1)$. This result lets us determine the relative strength
of defects observed in SGM images.

The largest uncertainty in the absolute depletion surface potential
$V_{i}^{*}$ of a given defect is the tip radius of curvature $R$,
taken to be 20 nm. The capacitance ratio $\alpha$ can be calculated
for known tip geometries and varies weakly $\alpha = 1.0 -1.5$ for
tip radii from 10 nm to 100 nm, and tip-surface separations of 1 -
100 nm; we take $\alpha\sim 1.1$ as a representative value.  For the
strong defects (numbered 1 - 4) seen in Fig. 2(a) - (e), we find
depletion surface potentials of 220 meV, 460 meV, 680 meV, and 340
meV, respectively, with uncertainties from data fitting of 10 - 20
meV. The two strongest defects are close to the electrodes,
suggesting that the Schottky barriers at the contacts lead to
additional band bending that enhances the impact of nearby defects.

The measured depletion surface potential can be converted into an
estimate of the local Fermi energy near the defect with $V_{tip}=0$.
As carriers are added to the depleted region, both the electrostatic
potential, $\Phi$, and the Fermi energy, $\epsilon_{F}$, increase.
With $N$ carriers added to a region of length $l$, these become
$e\Phi =Ne^{2}/Cl$, and $\epsilon_{F}=N/Dl$, where $C$ is the total
capacitance of the region per unit length, and $D$ is the density of
states per unit length, assumed constant for clarity. We estimate the
capacitance from a co-axial geometry ($C\sim
2\pi\epsilon\epsilon_{0}/ln(2h/r)\sim$ 30 pF/m = 0.2 e/V-nm, where
$h$ = 200 nm is the oxide thickness and $r$ = 1 nm the CNFET
radius\cite{Martel1}) and compare it to the ``quantum
capacitance"\cite{Lury} $e^{2}D$ = 2.4 e/V-nm, assuming four-fold
level degeneracy (spin and subband) and Fermi velocity $\hbar v_{F}$
= 0.5 eV-nm, appropriate for metallic nanotubes. Semiconducting tubes
have a larger density of states near the band edge but approach this
value away from the van Hove singularity. The ratio $C/e^{2}D = 1/12$
leads to an estimated Fermi energy of 20, 35, 50, and 25 meV for the
strong defects 1 - 4 in Fig. 2(a) - (e).

To further characterize defect related transport properties of the
nanotube, we perform simultaneous SIM (Ref. \cite{Kalinin}; schematic
Fig.\ 1(a)) to determine the local potential along the CNFET length.
The tip and backgate are held at constant voltages $V_{tip}$ and
$V_{g}$. A lateral bias: $V_{lat}=V_{dc}+V_{ac}cos(\omega t)$ is
applied to the sample to give an oscillating surface potential
$V_{surf}=V_{s}+V_{ac}(x)cos((\omega t)+\varphi(x))$, where
$V_{ac}(x)$ and $\varphi(x)$ are the amplitude and phase shift of the
voltage oscillation, and $V_{s}$ the dc surface potential. The first
harmonic of the electrostatic force on the tip is directly
proportional to $V_{ac}(x)$, and the amplitude and phase shift signal
from the cantilever reflect the amplitude and phase of the surface
potential oscillation.\cite{Kalinin}

When biased at a small negative voltage (e.g. $V_{tip}=$ -1 V), the
tip slightly enhances the local carrier density but acts as a
weakly-invasive probe of the potential distribution. In Fig. 4(a) we
see that when the CNFET is ``off" (poorly conducting, transport
current 40 nA) at $V_{g}=$ +0.7 V, SIM shows clearly resolved
potential steps at defects 2 and 4, while a series of weaker
scattering centers give a uniform voltage increase in the
neighborhood of defect 3. The nanotube is divided into multiple
conducting segments separated by distinct barriers at stronger
defects. Figure 4(b) shows that SIM indicates a {\it uniform
potential drop} along the device when the CNFET is ``on" (conducting,
transport current 350 nA) with $V_{g}=$ 0 V. All defects are far from
depletion, and each acts as a weak scatterering site: transport along
the CNFET is diffusive. We have reported evidence for ballistic
transport in semiconducting nanotubes at very high electrostatic
doping.\cite{Radosavljevic,CJ} SIM measurements in this regime are in
progress and will be reported elsewhere.

Figures 2(d)-(e) show that the large SGM signal from defects 3 and 4
obscure a set of weak defects that lie between these two. These
weaker defects can be studied using a ``tip-gated" SIM mode with a
large positive voltage applied to the tip gate. At this voltage, the
tip gate strongly alters the density and voltage profile beneath it.
If the tip voltage is sufficient to deplete a defect, a potential
step is created at the defect site, resulting in a large SIM signal
when the tip is precisely above the defect, with a sharp reduction in
signal as the tip moves away. This ``tip-gating" mode enables high
resolution imaging of weak defects, even if they lie near other
stronger defects (Fig.\ 2 (i)-(j)). This approach is extremely
effective in detecting weak defects and will be used in future
experiments to determine their associated Fermi energies.

In conclusion, we have studied defects in semiconducting SWNTs by
AC-SGM and SIM. The imaged defect size in SGM increases linearly with
tip potential, in agreement with an analysis of the electrostatic
interactions. This is used to determine the depletion surface
potential and Fermi energy for individual defects, ranging from 110 -
340 meV and 20 - 55 meV respectively, for the strong defects observed
in this experiment. SIM shows a series of potential drops located at
defect sites when the CNFET is in the ``off" state, and a uniform
potential drop along the CNFET, indicative of diffusive conduction,
when it is in the ``on" state. Finally, a novel self-gating SIM mode
is used to image defects with enhanced resolution even when they are
weak and located close to strong defects.

We acknowledge the support from MRSEC grant NSF DMR 00-79909 and
valuable discussions with M. Cohen, E.J. Mele, and  M.
Radosavljevi\'{c}.




\newpage

\begin{center}

{\bf Figure Captions \\[24pt]}

\end{center}

{\bf Figure 1. (a)} Setup for scanning gate microscopy (SGM) and
scanning impedance microscopy (SIM). In SGM, transport current is
measured as a function of tip-gate position. In SIM, the local
voltage oscillation is measured. {\bf(b)} AFM topography image of
the CNFET. {\bf(c)} Schematic of valence band energy modulation
due to defects. Vertical black bar is the local Fermi energy. The
tip depletes the shaded part of the tube below it. \\[12pt]

{\bf Figure 2. (a) - (e)} SGM images with tip voltage of 1, 2, 4,
6, and 8 V, respectively. Lateral bias is 0.3 $V_{pp}$ for (a) -
(c), 0.1 $V_{pp}$ for (d), (e). Back-gate voltage is -1 V. {\bf(f)
- (j)} Simultaneously acquired SIM images. Weak defects are
clearly resolved at high tip voltage (see text). Defects 1 - 4 and
the edges of the electrical contacts are shown in (c). The
diagonal line in (c) is
used for the SIM line scans of Fig. 4. \\[12pt]

{\bf Figure 3.} SGM imaged defect diameter increases linearly with
tip voltage. The slope of the line gives the depletion surface
potential of each defect. \\[12pt]

{\bf Figure 4. (a)} When the CNFET is in the ``off" state, SIM
shows voltage steps at the position of strong defects. The bias,
backgate, and tip voltages are $V_{ac} =$ 0.1 $V_{pp}$, $V_{g} =$
0.7 V, and $V_{tip} =$ -1 V. Lateral bias is $V_{ac} =$ 0.1
$V_{pp}$. At the voltage of $V_{tip} =$ -1 V the tip is a
non-invasive probe. {\bf(b)} SIM images a uniformly increasing
potential along the CNFET in the ``on" state (backgate = 0 V). The
line for both scans is indicated in
Fig. 2(c). \\[12pt]


\end{document}